\documentclass[twocolumn,superscriptaddress,aps,pra]{revtex4-2}

\usepackage{color}
\usepackage{graphicx}
\usepackage{hyperref}
\usepackage{amsmath,amssymb}
\usepackage{epstopdf}
\usepackage{subcaption}

\makeatletter
\newcommand*{\rom}[1]{\expandafter\@slowromancap\romannumeral #1@}
\makeatother
\def\beq{\begin{equation}}
	\def\eeq{\end{equation}}
\def\bea{\begin{eqnarray}}
	\def\eea{\end{eqnarray}}

\begin{document}
	
	\title{Intrinsic Attractive and Repulsive Interactions: From Classical to Quantum Statistics in the Generalized Maxwell-Boltzmann Distribution}
	
	\author{Maryam Seifi}
	\email{m.seifi.j@uma.ac.ir}
	\affiliation{Department of Physics, University of Mohaghegh Ardabili, P.O. Box 179, Ardabil, Iran}
	
	\author{Zahra Ebadi}
	\email{z.ebadi@uma.ac.ir}
	\affiliation{Department of Physics, University of Mohaghegh Ardabili, P.O. Box 179, Ardabil, Iran}
	
	\author{Hamzeh Agahi}
	\affiliation{Department of Mathematics, Faculty of Basic Science, Babol Noshirvani University of Technology, Shariati Ave., Babol 47148-71167, Iran}
	
	\author{Hossein Mehri-Dehnavi}
	\affiliation{Department of Physics, Faculty of Basic Science, Babol Noshirvani University of Technology, Babol 47148-71167, Iran}
	
	\author{Hosein Mohammadzadeh}
	\email{mohammadzadeh@uma.ac.ir}
	\affiliation{Department of Physics, University of Mohaghegh Ardabili, P.O. Box 179, Ardabil, Iran}
	
	\begin{abstract}
		The thermodynamic parameter space is flat for an ideal classical gas with non-interacting particles. In contrast, for an ideal quantum Bose (Fermi) gas, the thermodynamic curvature is positive (negative), indicating intrinsic attractive (repulsive) interactions. We generalize the classical Maxwell-Boltzmann distribution by employing a generalized form of the exponential function, proposing the Mittag-Leffler Maxwell-Boltzmann distribution within the framework of superstatistics. We demonstrate that the generalization parameter, $\alpha$, quantifies the statistical interaction. When $\alpha = 1$, the distribution coincides with the standard classical Maxwell-Boltzmann distribution, where no statistical interaction is present. For $0 < \alpha < 1$ ($\alpha > 1$), the statistical interaction is repulsive (attractive), corresponding to a negative (positive) thermodynamic curvature of the system.
	\end{abstract}
	
	\maketitle

	%%%%%%%%%%%%%%%%%%%%%%%%%%%%%%%%%%%%%%%%%%%%%%%%%%%
	\section{Introduction}\label{}
	%%%%%%%%%%%%%%%%%%%%%%%%%%%%%%%%%%%%%%%%%%%%%%%%%%%%

The introduction of generalized distributions in statistical mechanics represents a crucial effort for accurately determining and predicting the behavior of complex and unconventional systems. The initial generalization involves transitioning from classical distributions to quantum distributions within the framework of statistical mechanics. It is important to clarify that "distribution" refers to statistical distribution. Classical distributions, such as the Maxwell-Boltzmann (MB) distribution, are appropriate for distinguishable particles and situations where quantum effects are absent. Nevertheless, when considering quantum systems that involve identical particles ~\cite{leinaas1977theory}, such as electrons, it becomes essential to account for their indistinguishability and adherence to principles like the Pauli Exclusion Principle. In these cases, quantum distributions, such as the Fermi-Dirac (FD) and Bose-Einstein (BE) distributions, are essential for accurately describing these systems' behavior. These distributions consider the quantum nature of particles and provide a more precise representation of their behavior. Specifically, BE distribution enables the simultaneous occupation of the same quantum state by multiple particles, giving rise to phenomena like Bose-Einstein condensation (BEC) ~\cite{griffin1996bose}. In contrast, The FD distribution considers the exclusion principle, which prohibits fermions from occupying the same quantum state simultaneously ~\cite{shankar2012principles,pauli1940connection}.

These conventional distributions apply to ideal gas systems; however, many physical systems' inherent instability and complexity preclude them from adhering to the ideal gas model. By incorporating generalized distributions within the framework of statistical mechanics, a more comprehensive understanding of these complex systems can be achieved, leading to more accurate predictions of their behavior. This approach helps to overcome the constraints of traditional statistical models.

Intermediate distributions (Intermediate statistics) play a crucial role in explaining systems that do not adhere to conventional distributions \cite{haldane1991fractional,murthy1994haldane,polychronakos1996probabilities,khare2005fractional,green1953generalized,lee1990q,esmaili2024thermodynamic}. An example of intermediate distributions is fractional distributions, which has become evident through the observation of fractional resistances in the quantum Hall effect ~\cite{tsui1982two,canright1990fractional}. This discovery has identified particles known as Anyons ~\cite{wilczek1982magnetic,leinaas1977theory,bartolomei2020fractional}. Anyons display fractional statistics  In two-dimensional systems, exhibiting behavior that ranges between BE and FD distributions  ~\cite{leinaas1977theory}.

In systems characterized by long-range interactions, such as gravitational or Coulomb systems, generalized distributions like the Tsallis distribution are employed. These distributions encompass non-extensive entropy measures to effectively address the complexities of these interactions~\cite{jiulin2007nonextensivity,cohen2002statistics,plastino2004tsallis,tamarit2005relaxation,tsallis2002mixing}.

Superstatistics is an advanced concept within statistical mechanics that extends conventional distributions to encompass a broader array of systems. By defining superstatistics, it is possible to predict the behavior of complex, non-equilibrium systems characterized by fluctuations in intensive quantities, such as temperature, across spatial dimensions ~\cite{beck2003superstatistics}.

The Tsallis distribution and superstatistics are generalizations in statistical physics that extend the familiar Boltzmann factor $(e^{x} )$, $x=\beta\epsilon $ where $\epsilon$  represents the energy levels and $\beta=1/ k_{b} T$ with $T$ as the temperature and $k_{b}$ as the Boltzmann constant. The Tsallis distribution incorporates the control parameter $q$ to generalizations of the Boltzmann factor \cite{tsallis1988possible}. Likewise, Superstatistics offers a broader framework by accommodating systems that deviate from equilibrium, exhibiting fluctuations in intensive parameters (such as temperature or pressure) across different spatial regions or temporal intervals.  Superstatistics modifies the conventional Boltzmann factor to accommodate such fluctuations. Rather than relying on a singular \(\beta\), the system is characterized by a probability distribution function \(f(\beta)\), representing the spectrum of \(\beta\) values and thereby addressing the varying intensive parameters. The generalized Boltzmann factor in superstatistics is expressed as follows \cite{beck2003superstatistics}:

$$\mathnormal{B(\epsilon)}=\int_{0}^{\infty} d\beta \mathnormal{f}(\beta) \mathnormal{e} ^{-\beta \epsilon} .$$

 In addition, Kaniadakis ~\cite{kaniadakis2021new} and Dunkl ~\cite{dunkl1989differential} introduce novel distributions by generalizing the Boltzmann factor. Remarkably, mathematically, these generalizations extend the standard exponential function, offering a broader framework for understanding complex systems \cite{cheraghalizadeh2021superstatistical}.

The Mittag-Leffler (ML) function  ~\cite{gorenflo2020mittag} is a powerful mathematical tool with many applications. Its ability to generalize the exponential function makes it particularly useful in various fields of science and engineering. Its significance has progressively emerged within the domain of physics ~\cite{dos2020mittag,mathai2010mittag}.
 Nevertheless, despite its potential importance, physicists have yet to explore its applications, leaving many opportunities for experts in these fields to delve deeper into its properties and potential applications. The function, named after the renowned Swedish mathematician Gösta Mittag-Leffler, provides a sturdy mathematical framework for understanding complex, non-exponential behaviors in various physical systems ~\cite{dattoli2017comments,agahi2019mittag,agahi2017pseudo}.

 The current research aims to generalize MB distribution by incorporating the ML function. This incorporation attempts to account for more complex interactions or behaviors that the classical distribution may not fully capture. By doing so, the researchers intend to provide a more comprehensive analysis of the thermodynamic properties associated with the system under consideration. This could lead to a better understanding of the system's behavior under different conditions, potentially yielding insights into real-world phenomena where classical assumptions may not hold.

 Different techniques are available in statistical mechanics to study and analyze distributions in various conditions. These methods serve as invaluable tools for elucidating the intricate dynamics governing particle interactions within generalized distributions. By employing these methods, scientists can deepen their understanding of how particles behave under different circumstances and distributions. Among these methods, geometric thermodynamics, with its systematic approach, stands out as a commonly used method for studying particle interactions.

Gibbs was a pioneer in exploring the geometric aspects of thermodynamic systems. Subsequently, Ruppeiner and Weinhold established the groundwork for the advancement of thermodynamic geometry ~\cite{gibbs1928thermodynamics,
ruppeiner1979thermodynamics,
weinhold1975metric}. Ruppeiner's formalism, based on the fluctuation theory, defined a thermodynamic parameter space and an associated metric, derived from the second derivatives of entropy with respect to intrinsic extensive parameters such as internal energy, volume, and particle number. Similarly, the Weinhold metric is derived from the second derivatives of internal energy with respect to the related extensive parameters. It has been shown that the Ruppeiner and Weinhold metrics are conformally equivalent ~\cite{ruppeiner1995riemannian,
salamon1984relation}.

The thermodynamic curvature, often called the Ricci scalar, is a key quantity derived from the metric and is pivotal in analyzing thermodynamic systems. For example, the thermodynamic curvature of a single-component ideal gas is zero. Janyszek and Mrugała's investigations into the curvature of ideal quantum gases revealed that an ideal Bose gas consistently has a positive curvature. In contrast, an ideal Fermi gas has a negative curvature. Although arbitrary, the sign of the thermodynamic curvature categorizes the intrinsic statistical interactions among particles ~\cite{ruppeiner1995riemannian,janyszek1990riemannian,oshima1999riemann}.
 Thermodynamic curvature shows singular behavior at phase transition points, such as the condensation point in various boson gases~\cite{janyszek1990riemannian,mirza2011thermodynamic,adli2019condensation,mohammadzadeh2016perturbative}.

Thermodynamic geometry is valid for analyzing systems with equilibrium dynamics.  However, it has also been effectively utilized in certain nonequilibrium systems that conform to generalized statistical distributions rather than conventional ones, such as the Maxwell-Boltzmann distribution or quantum distributions like Bose-Einstein and Fermi-Dirac.
 Notable examples include Tsallis statistics, Kaniadakis statistics, and q-deformations, which have been successfully applied to various physical phenomena, such as plasmas, finite-size systems, anomalous diffusion in ultracold atoms, harmonic spring-mass systems, scalar mixing in the interstellar medium, and colloidal particles in the cytoplasm ~\cite{sagi2012observation,saporta2019self,colbrook2017scaling}. 
In general, for systems with complex interactions and non-equilibrium conditions, generalized statistics serves as an effective approach, since it practically corresponds to equilibrium systems, albeit with some additional parameters. From this perspective, the calculation of thermodynamic geometry in non-equilibrium statistics remains a relevant and valid exercise.

The thermodynamic geometry has been widely employed to investigate the thermodynamic properties of various statistical systems, emphasizing the effect of a discrete ground state on the thermodynamic curvature of the system~\cite{pessoa2021information,ubriaco2012scalar}. Moreover, the thermodynamic geometry of intermediate statistics~\cite{mirza2009nonperturbative,mirza2010thermodynamic,oshima1999riemann}, fractal distributions~\cite{lopez2021information}, non-extensive statistics~\cite{mohammadzadeh2016perturbative,adli2019condensation}, deformed statistics~\cite{mirza2011thermodynamic}, and Kaniadakis statistics~\cite{mehri2020thermodynamic} has been extensively investigated.

In Sec. ($\bold{II}$), we delineate the procedure for creating the generalized distribution, elucidating its connection to both MB distribution and ML functions. Sec. ($\bold{III}$)   is dedicated to an in-depth examination of thermodynamic geometry and a brief discussion on the computation of thermodynamic curvature. ($\bold{V}$)  we analyze heat capacity and comprehensively explore the condensation phenomenon in the context of quantum statistics. Lastly, in Sec. ($\bold{VI}$), we present concluding remarks to summarize the key findings of this paper.

\section{ Mittag-Lefler Function As a Generalized Distribution}\label{}
%%%%%%%%%%%%%%%%%%%%%%%%%

\begin{figure}[t]
    \includegraphics[scale=0.5]{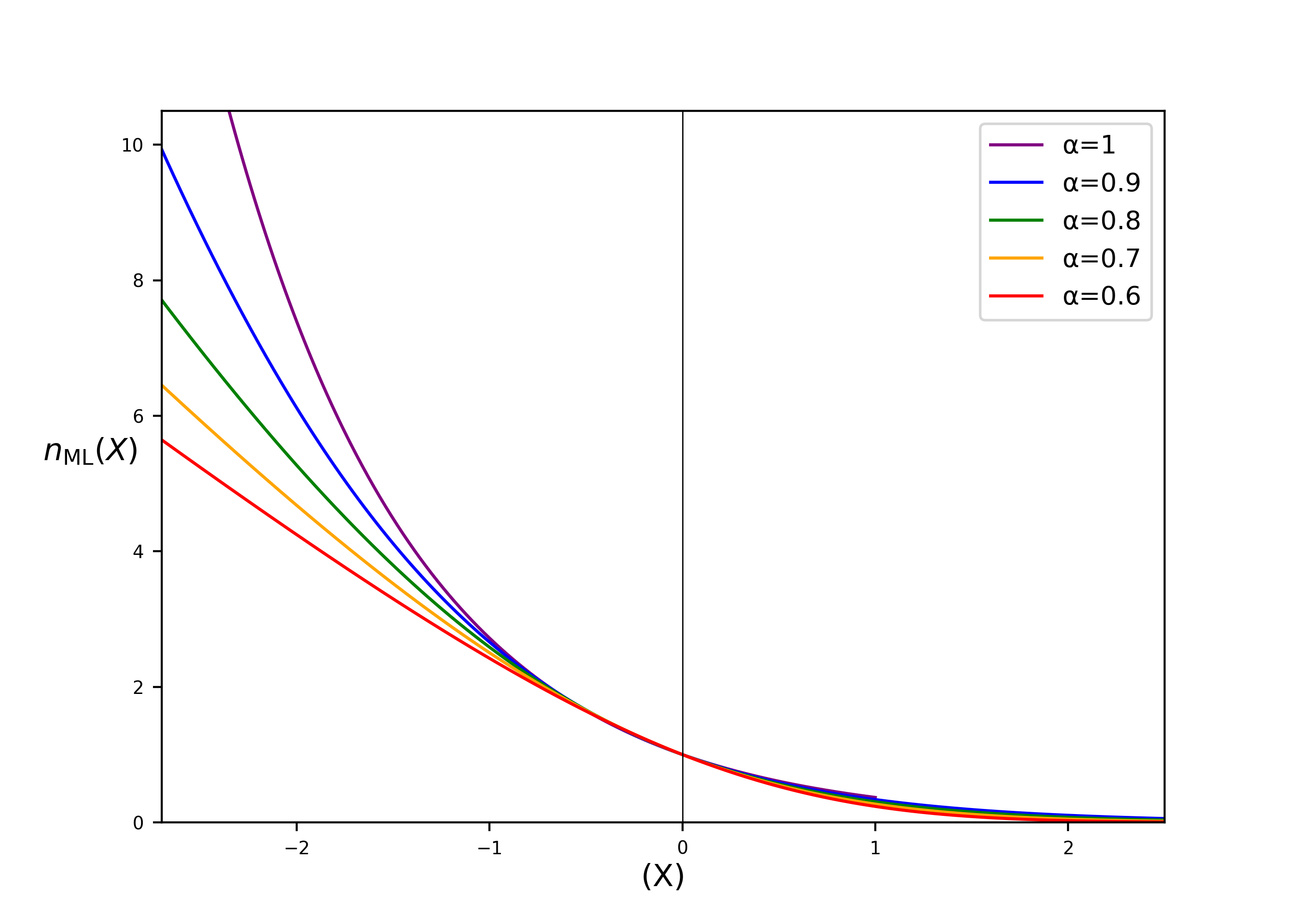}

    \captionsetup{justification=raggedright, singlelinecheck=false} % Left-align caption
    \caption{%
        \( n_{\text{ML}}(X) \) of the MLMB distribution as a function of \(X\), plotted for the range \(0 \leq \alpha \leq 1\). 
        The solid purple line represents the case \(\alpha = 1\) (MB distribution), while the other lines correspond to \(\alpha = 0.6, 0.7, 0.8, 0.9\).
    }
    \label{ne a kam1}
\end{figure}

In statistical mechanics, distributions are crucial for describing the states of physical systems. One well-known example is MB distribution, which applies to classical particles. It statistically portrays how these particles are distributed among different energy states when the system reaches thermal equilibrium. At the quantum level, two main distributions play a role: BE and FD distributions. These distributions provide valuable insights into the distinctive characteristics of bosons and fermions, two fundamental classifications of quantum particles.

While these distributions are typically used for ideal equilibrium systems, the complexity and non-equilibrium nature of many real-world systems require generalization. Researchers in statistical mechanics delve into this area to develop a more comprehensive understanding of physical system behavior beyond idealized models. The study of generalized statistical distributions reflects the complexities inherent in physical systems. Generalizing conventional statistical distributions is a crucial aspect of statistical mechanics, allowing for analyzing systems beyond the scope of standard models.

One such approach involves generalized intermediate statistics, exemplified by Haldane \cite{haldane1991fractional,murthy1994haldane}, Polychronakos ~\cite{polychronakos1996probabilities}, Gentil ~\cite{khare2005fractional} and q-deformed \cite{green1953generalized,lee1990q} statistics. Several generalizations pertain to the concept of generalized exponential functions, including Tsallis \cite{tsallis1994numbers,tsallis2009nonadditive,umarov2008aq} and Kanadiakis statistics \cite{kaniadakis2001non,kaniadakis2021new}. Also,  in \cite{beck2003superstatistics}  a significant role in developing the generalized Boltzmann factor was investigated. In addition, a multi-parameters generalized exponential function called ML function has received considerable attention in the literature ~\cite{gorenflo2020mittag}.

\begin{figure}[t]
    \includegraphics[scale=0.5]{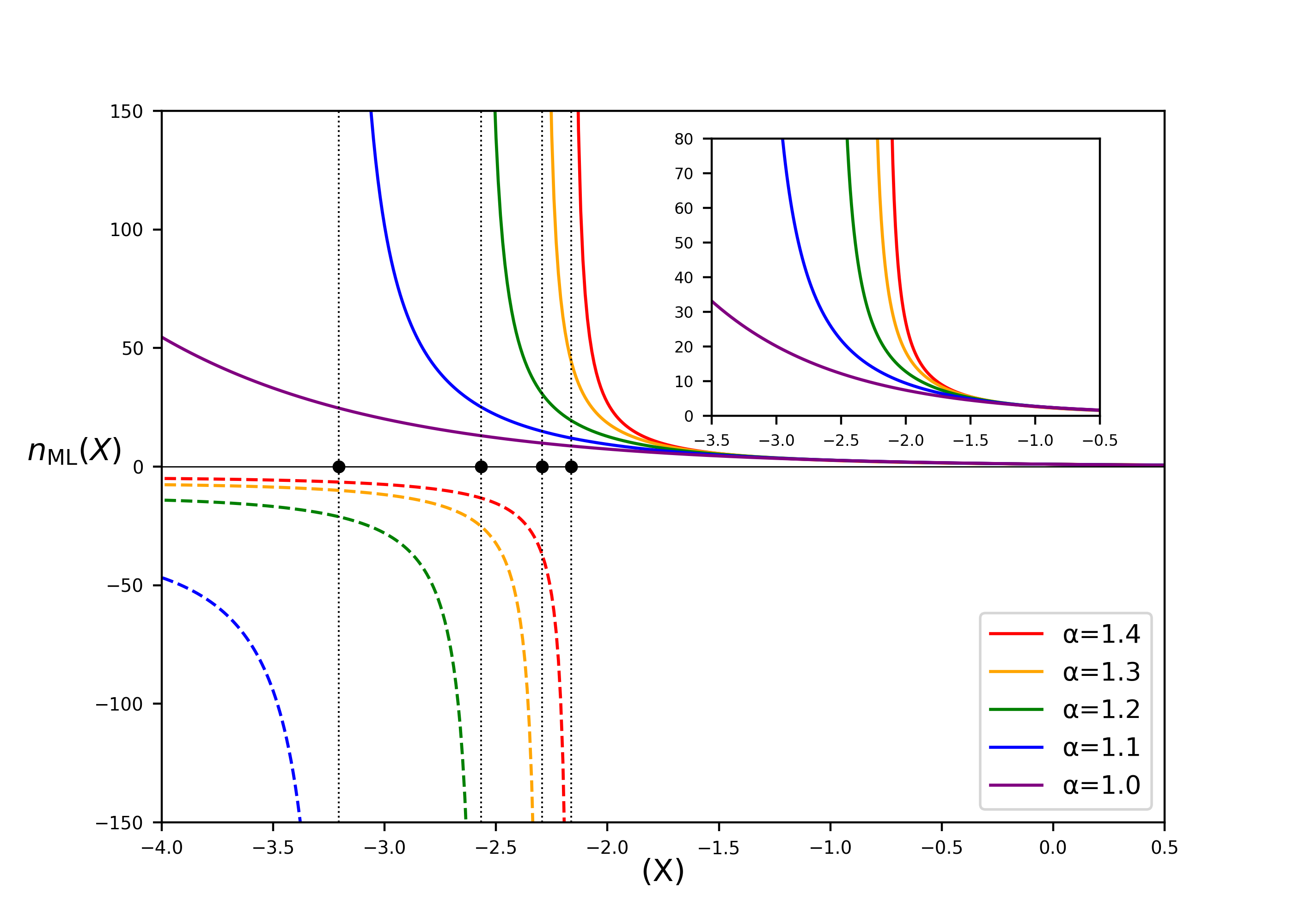}

    \captionsetup{justification=raggedright, singlelinecheck=false} % Left-align caption
    \caption{%
        \( n_{\text{ML}}(X) \) of the MLMB distribution as a function of $X$, plotted for the range \( 1 \leq \alpha \leq 1.5 \). 
        Solid lines indicate positive values of \( n_{\text{ML}}(X) \) for \( X > X_{\mathrm{th}}^{\alpha} \), 
        while dashed lines denote negative values for \( X < X_{\mathrm{th}}^{\alpha} \). 
        Circular markers represent the threshold points \( X_{\mathrm{th}}^{\alpha} \) for each \( \alpha \), 
        where the sign of \( n_{\text{ML}}(X) \) changes. 
        The purple solid line corresponds to \( \alpha = 1 \), which represents the MB distribution. 
        The inset provides a detailed view near the threshold points, emphasizing the distinct behavior of 
        \( n_{\text{ML}}(X) \) for \( \alpha = 1.1, 1.2, 1.3, 1.4 \).
    }
    \label{ne_a_bish_1(+,-)}
\end{figure}

The ML function, highly esteemed by researchers in applied sciences and various engineering disciplines, plays a crucial role in several areas, including stochastic systems, dynamical systems theory, and disordered systems \cite{kilbas2006theory,podlubny1998fractional,gorenflo1997fractional,mainardi2000mittag,hilfer2000applications,haubold2011mittag}. 
The ML function is a complex function with many applications in fractional calculus. Specifically, they describe deviations of physical phenomena from exponential behavior. A power series delineates this function and entails two complex parameters: $\alpha$ and $\lambda$. {It becomes an entire function when $\alpha$ and $\lambda$ are positive and real.} Analogous to the exponential function that arises from the solution of ordinary differential equations, the ML function plays a similar role in the solution of fractional differential equations. The exponential function is a particular instance of the ML function. The power series representation of the exponential function can be expressed as follows:
\begin{equation}
\exp(X) = \sum_{k=0}^{\infty} \frac{X^{k}}{\Gamma[k+1]},
\end{equation}
where $\Gamma$ denotes the Gamma function. This can be generalized to the ML function as follows:
\begin{equation}
    E_{\alpha,\lambda}(X) = \sum_{k=0}^{\infty} \frac{X^{k}}{\Gamma[\alpha k + \lambda]},
\end{equation}
Which reduces to the exponential function when $\alpha = \lambda = 1$~\cite{haubold2011mittag,camargo2012generalized,gorenflo2020mittag,kurulay2012some,gorenflo1997fractional}. In our research, we strive to develop an innovative distribution function that can accurately reproduce the MB distribution by utilizing the ML function. The foundation for achieving this goal involves replacing the standard exponential function in the MB distribution with the ML function.

Our investigation focuses on the single-parameter the ML function with $\lambda = 1$, which can accommodate any positive value of $\alpha$ as follows:
\begin{equation}
E_{\alpha,1}(X) = E_{\alpha}(X) = \sum_{k=0}^{\infty} \frac{X^k}{\Gamma[\alpha k + 1]}.
\end{equation}

However, we narrow our attention to $0 < \alpha \le 1.5$ for this research. Accordingly, by incorporating the ML function into the occupation number distribution of MB distribution, we derive a novel function expressed in the following form. This function will be referred to as the Mittag-Leffler Maxwell-Boltzmann distribution (MLMB distribution):
\bea
n_{\mathrm{ML}}(X)&=\frac{1}{E_{\alpha}(X)},
\label{n_{ ML-MB}}
\eea
where $n_{\mathrm{ML}}(X)$ is the occupation number, $X=(x-\beta\mu)$ while $\mu$ is the chemical potential.  The fugacity is defined by $ z=e^{\beta\mu}=e^{- \gamma}$ and therefore, $X=(x - \ln \mathnormal{z})$.

The requirement for positive definiteness for distributions is significant as it places constraints on physical quantities, thereby preventing unphysical negative occupation numbers. This requirement ensures that the distribution remains within the realm of physically plausible scenarios. The positivity condition of the BE distribution necessitates the fugacity to lie within the interval \(0 \leq z < 1\). Accordingly, the maximum permissible value for the fugacity is \(z = 1\). In the subsequent discussion, we will investigate \( n_{\mathrm{ML}}(X) \) on the generalized MLMB distribution.

Fig. \ref{ne a kam1} illustrates the relationship between \( n_{\mathrm{ML}}(X) \) and \( X \) is presented for the interval \( 0 < \alpha \leq 1 \). Within this range, \( n_{\mathrm{ML}}(X) \) consistently remains positive across all values of \( X \). On the other hand, Fig. \ref{ne_a_bish_1(+,-)} depicts  the relationship between \( n_{\mathrm{ML}}(X) \) and \( X \) for \( 1 \leq \alpha < 1.5 \). In this figure, circular markers along the horizontal axis represent specific threshold values, \( X_{\mathrm{th}}^{\alpha} \), corresponding to each value of \( \alpha \). These threshold values divide the graph into two regions: one where \( n_{\mathrm{ML}}(X) \) is positive for \( X > X_{\mathrm{th}}^{\alpha} \), indicating a permissible range for the occupation number, and another where \( n_{\mathrm{ML}}(X) \) becomes negative for \( X < X_{\mathrm{th}}^{\alpha} \), indicating an impermissible range. Therefore, to ensure the positivity of \( n_{\mathrm{ML}}(X) \), it is necessary to establish a minimum value for \( X \) such that \( n_{\mathrm{ML}}(X) \) remains positive for all \( X > X_{\mathrm{th}}^{\alpha} \).

\begin{figure}[t]
    \includegraphics[scale=0.5]{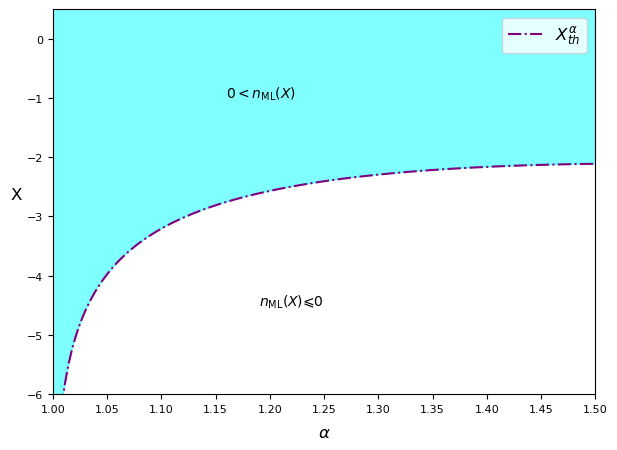}

    \captionsetup{justification=raggedright, singlelinecheck=false} % Left-align caption
    \caption{%
        Variation of \( X \) as a function of \( 1 < \alpha \leq 1.5 \). 
        The dashed purple line represents the values of \( X_{\mathrm{th}}^{\alpha} \). 
        The cyan regions above the dashed line indicate where \( n_{\mathrm{ML}}(X) \) is positive, 
        while the white regions below the dashed line indicate where \( n_{\mathrm{ML}}(X) \) is negative.
    }
    \label{mroot}
\end{figure}

The relationship between \( X_{\mathrm{th}}^{\alpha} \) and \( \alpha \) is depicted in Fig. ~\ref{mroot}. Areas where \( X > X_{\mathrm{th}}^{\alpha} \) correspond to positive values of \( n_{\mathrm{ML}}(X) \), while areas where \( X < X_{\mathrm{th}}^{\alpha} \) are associated with negative values of \( n_{\mathrm{ML}}(X) \).

This analysis attempts to ascertain the constraints on the values of \( z \) for varying values of \( \alpha \). As previously noted, when \( X = (\beta \epsilon - \ln z) \geq X_{\mathrm{th}}^{\alpha}  \), the function \( n_{\mathrm{ML}}(X) \) remains positive. In this case, \( \beta \) represents the inverse of the thermodynamic temperature and is always positive. Considering that the energy \( \epsilon \) cannot be negative, the non-negativity of \( n_{\mathrm{ML}}(X) \) is contingent upon the maximum value of \( z \) (\( z_{\mathrm{th}}^{\alpha} \)), such that \( -\ln (z_{\mathrm{th}}^{\alpha}) = X_{\mathrm{th}}^{\alpha}  \),  when \( \epsilon \) is set to zero and therefor, \(  z_{\mathrm{th}}^{\alpha} =( e^{-X_{\mathrm{th}}^{\alpha} } \)). Accordingly, for the MLMB distribution when \( 1 < \alpha \leq 1.5\), the constraint on \( z \) is \( 0 \leq z \leq  z_{c}^{\alpha} \). 

\begin{figure}[t]
    \includegraphics[scale=0.5]{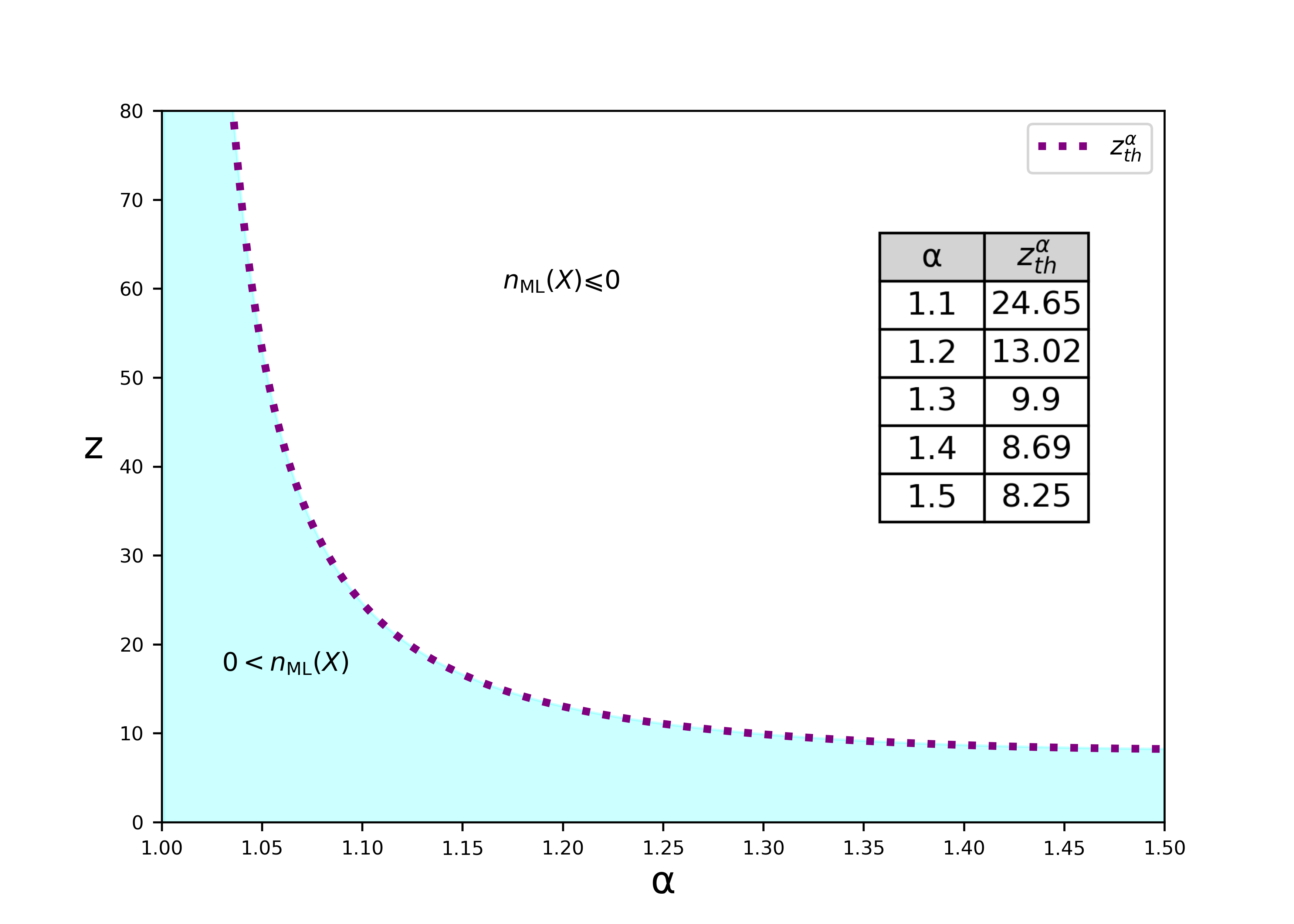}

    \captionsetup{justification=raggedright, singlelinecheck=false} % Left-align caption
    \caption{%
        Variation of \( z \) with respect to \( 1 < \alpha \leq 1.5 \). 
        The dashed purple line represents \( z_{\mathrm{th}}^{\alpha} \) values. The cyan areas (below the dashed line) indicate where \( n_{\mathrm{ML}}(X) \) is positive, while the white areas (above the dashed line) indicate where \( n_{\mathrm{ML}}(X) \) is negative.
    }
    \label{z_max_be_ezay_alpha_ha_plot}
\end{figure}

Fig. ~\ref{z_max_be_ezay_alpha_ha_plot} illustrates the relationship between $z$ and the parameter $\alpha$. It is shown that $n_{\mathrm{ML}}(X)$ remains non-negative when $z <  z_{\mathrm{th}}^{\alpha}$. The bounded allowed values for fugacity for $1<\alpha\le1.5$ are similar to the bounded values for fugacity of the bosons in BE distribution. Accordingly, A similar phenomenon is expected at $z= z_{\mathrm{th}}^{\alpha}$ in the MLMB distribution for $1<\alpha\le1.5$.

\subsection{Thermodynamic Quantities of MLMB Distribution }

The crucial thermodynamic functions for the subsequent discussion are the internal energy and the total particle number. In the thermodynamic limit, for non-interacting particles described by the distribution function \(n(\epsilon)\) and confined within a \(D\)-dimensional box of volume \(L^D\), the dispersion relation is given by \(\epsilon = a p^\sigma\), where \(a\) is a constant, \(p\) denotes momentum, and \(\sigma\) is an exponent defining the relationship between energy and momentum. When \(\sigma = 2\), this relation describes non-relativistic particles and simplifies to \(\epsilon=a p^2\). This form is analogous to the kinetic energy expression \(\epsilon = p^2/2m\) for a free particle, with \(a=1/2 m\), where \(m\) represents the particle's mass. Accordingly, for such particles, the expressions for the internal energy and total number of particles can be written as follows:

\begin{equation}
	\begin{split}
		&U = \int _{0}^{\infty} \epsilon n(\epsilon) \Omega(\epsilon) d\epsilon,  \\
		&N = \int _{0}^{\infty}  n(\epsilon) \Omega(\epsilon) d\epsilon.   \\ 
	\end{split}
	\label{uN mb}
\end{equation}

The single-particle density of states for the system is given by:
\begin{equation}
\Omega(\epsilon)=A^{D}\epsilon^{\frac{D}{\sigma}-1},
      \label{OMEGA}
\end{equation}
where the parameter \( A \) is defined as:
  \begin{equation}
    A=\frac{L\sqrt{\pi}}{(\Gamma[\frac{D}{2}])^{\frac{1}{D}}a^{\frac{1}{\sigma}}h},
      \label{OMEGA2}
\end{equation}

Where \( D \) represents the dimensionality of the system. For the specific case of a $3$-dimensional system with non-relativistic particles, the expression reduces to \(\Omega(\epsilon) = \epsilon^{1/2}\), with \( A \) set to unity for simplicity.

Using Eqs. (\ref{n_{ ML-MB}}) and (\ref{uN mb}), the functions \(U\) and \(N\)  within the MLMB distribution can be expressed as follows:

\begin{equation}
	\begin{split}
		& U = \beta^{-\frac{5}{2}}\int _{0}^{\infty} \frac{x^{\frac{3}{2}}}{\mathnormal{E}_{\alpha}(\mathnormal{x - \ln \mathnormal{z}})} dx, \\
		&\mathnormal{N}= \beta^{-\frac{3}{2}}\int _{0}^{\infty} \frac{x^{\frac{1}{2}}}{\mathnormal{E}_{\alpha}(\mathnormal{x - \ln \mathnormal{z}})} dx.  \\
	\end{split}
	\label{uN mlmb}
\end{equation}

To enhance the clarity and appearance of the relationships, we can utilize the following notation:

\bea
	\int _{0}^{\infty} \frac{ x^{n}}{{E}_{\alpha}({x - \ln \mathnormal{z}})} dx = \mathcal{H}_{n,\alpha}(z)
\label{nnn_{ ML-MB}}
\eea

Therefore, we can rephrase the previous relationship in the following manner:

\begin{equation}
	\begin{split}
		& U =  \beta^{-\frac{5}{2}} \mathcal{H}_{\frac{3}{2},\alpha}(z) , \\
		&\mathnormal{N}=  \beta^{-\frac{3}{2}} \mathcal{H}_{\frac{1}{2},\alpha}(z) .  \\
	\end{split}
	\label{uNgg mlmb}
\end{equation}

These refined formulations simplify the expressions while establishing a more coherent framework for further analytical or numerical investigations into the MLMB distribution.

\section{Thermodynamic Curvature of the MLMB Distribution} \label{}

Ruppeiner and Weinhold made significant contributions to the field of thermodynamic geometry\cite{ruppeiner1979thermodynamics, weinhold1975metric}. In this framework, the parameter space of thermodynamic variables is construed as a Riemannian manifold, facilitating the formulation of an appropriate metric tensor within this space. The Ruppeiner metric is established through the computation of second-order derivatives of entropy with respect to the relevant extensive thermodynamic variables, including internal energy, volume, and the total number of particles. Additionally, Weinhold introduced an alternative metric in the energy representation, defined by the second-order derivatives of internal energy concerning the pertinent extensive thermodynamic parameters. Application of a Legendre transformation to either entropy or internal energy with respect to the extensive parameters yields various thermodynamic potentials, such as the Helmholtz and Gibbs free energies. The Fisher-Rao metric is delineated by the second derivatives of the logarithm of the partition function with respect to the non-extensive thermodynamic parameters as follows
\cite{ruppeiner1995riemannian, janyszek1990riemannian}.
\bea
g_{ij} = \frac{\partial^2 \ln  \mathcal{Z}}{\partial \beta^i \partial \beta^j},
\label{fi}
\eea
where, ${\beta^{1}=\beta,\beta^{2}=}\gamma $ and $  \mathcal{Z} $ representing the partition function.% The metric components are derived through the subsequent relational expression:

\begin{figure}[t]
    \includegraphics[scale=0.5]{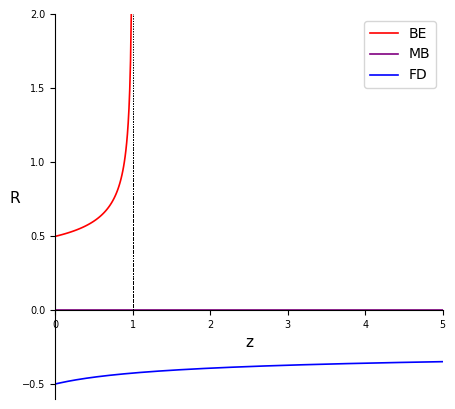}

    \captionsetup{justification=raggedright, singlelinecheck=false} % Left-align caption
    \caption{%
        Thermodynamic curvature as a function of fugacity for an ideal gas, comparing the Bose-Einstein distribution (red solid line) and the Fermi-Dirac distribution (blue solid line), under isothermal conditions (\(\beta = 1\)).
    }
    \label{R z mb fd be 1}
\end{figure}

The space of thermodynamic parameters is set to two when considering a system with two degrees of freedom. Janyszek and Mrugała have elucidated that if the metric components are exclusively expressed as the second derivatives of a specified thermodynamic potential, then the thermodynamic curvature can be articulated in terms of the second and third derivatives. Within the confines of two-dimensional spaces, the Ricci scalar is defined accordingly as follows~\cite{janyszek1989riemannian}

 \bea
 R = \frac{  \begin{vmatrix}			
 		g_{\beta \beta} & 	g_{\gamma \gamma} & 	g_{\beta \gamma} \\		
 		g_{\beta \beta, \beta} & 	g_{\gamma \gamma, \beta} & 	g_{\beta \gamma, \beta} \\
 		g_{\beta \beta,\gamma} & 	g_{\gamma \gamma, \gamma} & 	g_{\beta \gamma, \gamma} \\
 \end{vmatrix}}
 {2 {{\begin{vmatrix}
 				g_{\beta \beta} & 	g_{\beta \gamma} \\ 		
 				g_{\beta \gamma} & 	g_{\gamma \gamma} 			
 	\end{vmatrix}}}^{2}}.
 \label{rrr}
 \eea

 It has been shown that within the thermodynamic parameter space of an ideal gas with particles obeying the MB distribution, the curvature of the thermodynamic space is zero. Conversely, in ideal quantum gases, the thermodynamic curvature is positive for bosons and negative for fermions. This implies that the sign of thermodynamic curvature reflects the intrinsic statistical interactions within the system: positive curvature corresponds to intrinsic attractive interactions, while negative curvature corresponds to repulsive interactions~\cite{ruppeiner1995riemannian,janyszek1990riemannian,oshima1999riemann}. Furthermore, the behavior of thermodynamic curvature can offer insights into phase transitions, such as the singularity observed at the condensation point of an ideal boson gas~\cite{janyszek1990riemannian,mirza2011thermodynamic,adli2019condensation,mohammadzadeh2016perturbative}.

\begin{figure}[t]
    \includegraphics[scale=0.5]{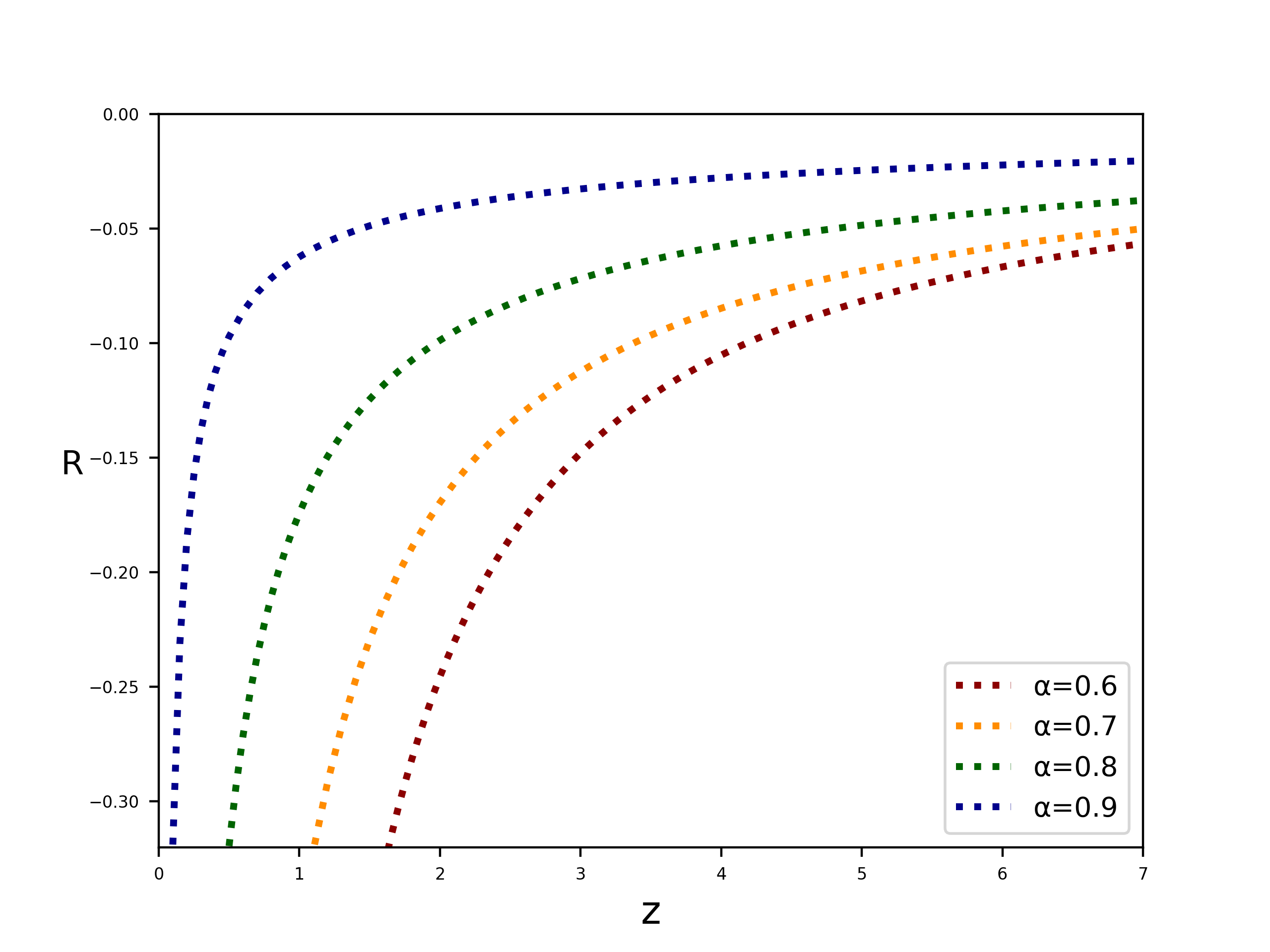}

    \captionsetup{justification=raggedright, singlelinecheck=false} % Left-align caption
    \caption{%
        Thermodynamic curvature of an MLMB distribution as a function of fugacity, plotted for the range \(0 < \alpha < 1\) under isothermal conditions (\(\beta = 1\)). 
        Dashed lines represent \(\alpha = 0.6, 0.7, 0.8, 0.9\).
    }
    \label{R z mb a kam}
\end{figure}

The thermodynamic curvature for an ideal boson gas and an ideal fermion gas in three dimensions, with non-relativistic particles, are illustrated in Fig. (\ref{R z mb fd be 1}). In the depicted graph, where thermodynamic curvature is plotted against $z$, it is evident that for the FD distribution, $R$ is negative for all values of $z$.  Conversely, in the BE distribution, $R$ is positive for all allowed values of $0\le z < 1$ and diverges at $z = 1$, indicating the onset of BEC. In the MB distribution case, $R$ remains zero regardless of $z$. In the subsequent sections, we will investigate the thermodynamic curvature in the context of the generalized MLMB distribution.

By referencing Eqs. (\ref{fi}) and substituting Eq. (\ref{uNgg mlmb}) therein, the metric components associated with the MLMB distribution can be determined as follows:

 \begin{equation}
	\begin{split}
		&g_{\beta \beta}  = \frac{\partial^2 \ln  \mathcal{Z}}{\partial \beta^2 }=-(\frac{\partial U}{\partial \beta })_{\gamma}= \frac{5}{2 } \beta^{-\frac{7}{2}} \mathcal{H}_{\frac{3}{2},\alpha}(z),
\\
		&g_{\beta \gamma}  = \!
 g_{\gamma \beta } = \frac{\partial^2 \ln  \mathcal{Z}}{\partial \beta \partial \gamma }=-(\frac{\partial N}{\partial \beta})_{\gamma}=\frac{3}{2 } \beta^{-\frac{5}{2}} \mathcal{H}_{\frac{1}{2},\alpha}(z),
  \\
		&g_{\gamma \gamma}  = \!
\frac{\partial^2 \ln  \mathcal{Z}}{ \partial \gamma^{2} }=-(\frac{\partial N}{\partial \gamma})_{\beta}= z \beta^{-\frac{3}{2}} \partial_{z} ( \mathcal{H}_{\frac{1}{2},\alpha}(z)).
 \\	
 	\end{split}
 	\label{g mlmb1}
 \end{equation}

\begin{figure}[t]
    \includegraphics[scale=0.5]{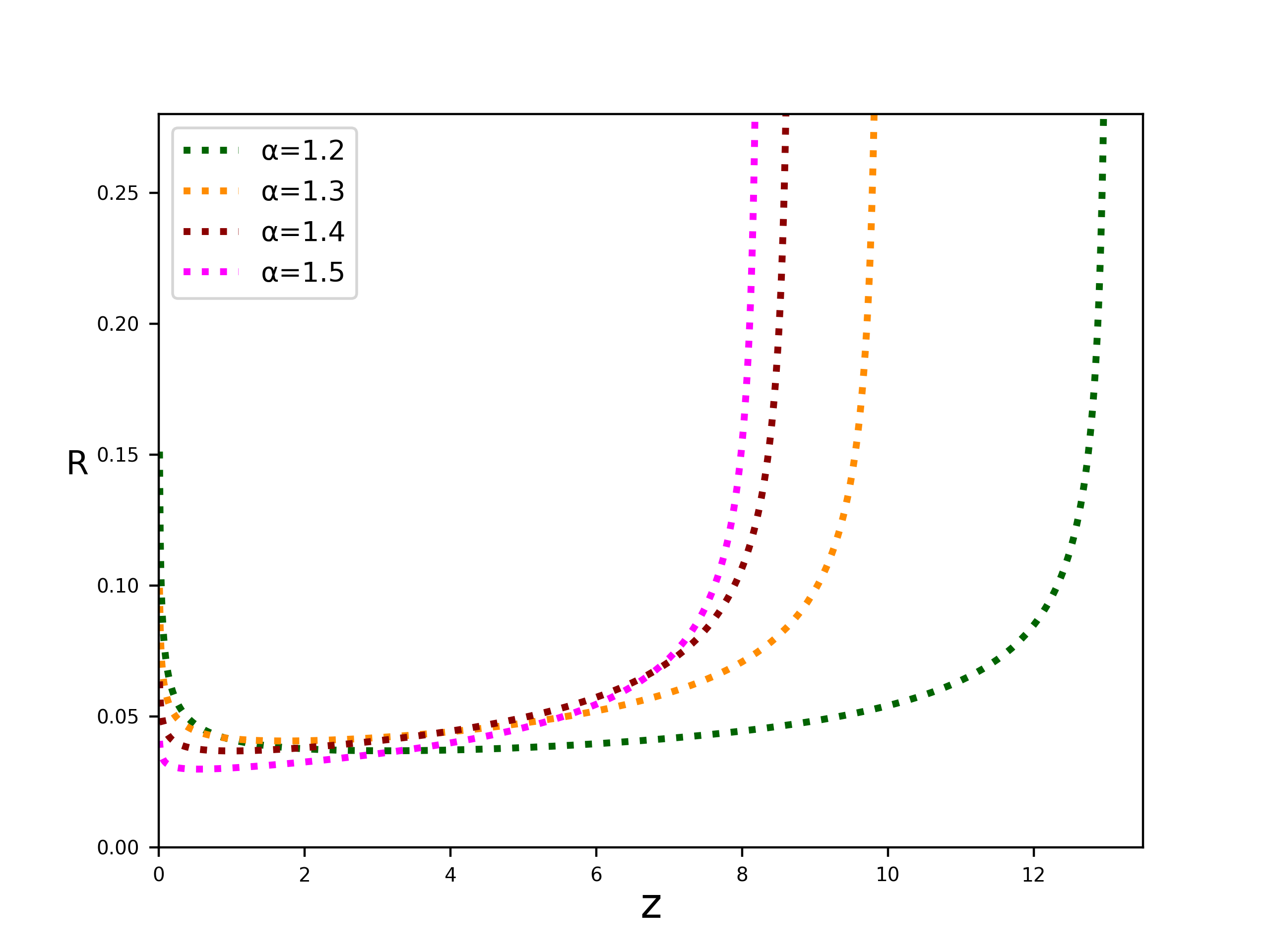}

    \captionsetup{justification=raggedright, singlelinecheck=false} % Left-align caption
    \caption{%
        Thermodynamic curvature of an MLMB distribution as a function of fugacity, plotted for the range \(1 < \alpha \leq 1.5\) under isothermal conditions (\(\beta = 1\)). 
        Dashed lines represent the values \(\alpha = 1.2, 1.3, 1.4, 1.5\).
    }
    \label{R z mb a bish}
\end{figure}

For evaluating the last equation, we use the appendix \ref{appendix}. By derivation from Eqs. (\ref{g mlmb1}) with respect to \(\gamma\) and \(\beta\), we can derive additional elements of Eqs. (\ref{rrr}) as follows:

\begin{equation}
\begin{aligned}
    g_{\beta \beta,\beta} &= \frac{\partial}{\partial \beta} g_{\beta \beta}= -\frac{35}{4} \beta^{-\frac{9}{2}} \mathcal{H}_{\frac{3}{2},\alpha}(z), \\
    \\
    g_{\beta \beta,\gamma} &= g_{\beta \gamma,\beta} = g_{\gamma \beta,\beta} = \frac{\partial}{\partial \gamma} g_{\beta \beta} = -\frac{5}{2} \beta^{-\frac{7}{2}} \mathcal{H}_{\frac{3}{2},\alpha}(z), \\
    \\
    g_{\beta \gamma,\gamma} &= g_{\gamma \beta,\gamma} = g_{\gamma \gamma,\beta} = \frac{\partial}{\partial \beta} g_{\gamma \gamma}= -\frac{3}{2} z \beta^{-\frac{5}{2}}  \partial_{z} \mathcal{H}_{\frac{1}{2},\alpha}(z), \\
    \\
    g_{\gamma \gamma,\gamma} &= \frac{\partial}{\partial \gamma} g_{\gamma \gamma}=-z ( \beta^{-\frac{3}{2}} \partial_{z} \mathcal{H}_{\frac{1}{2},\alpha}(z)  + z \beta^{-\frac{3}{2}} \partial^{2}_{z} \mathcal{H}_{\frac{1}{2},\alpha}(z)).
\end{aligned}
\label{g mlmb2}
\end{equation}

By integrating Eqs. (\ref{g mlmb1}) and (\ref{g mlmb2}) into Eqs. (\ref{rrr}), we can construct a graph analogous to Fig. (\ref{R z mb fd be 1}), specifically for the MLMB distribution, as illustrated in Figs. (\ref{R z mb a kam}) and (\ref{R z mb a bish}). These graphical representations are instrumental in facilitating a visual comparison and in-depth analysis of geometric thermodynamics. 

The diagrams highlight the crucial role of the parameter \(\alpha\) in determining the thermodynamic curvature of particles within the MLMB distribution.
For \(0 < \alpha < 1\), the system demonstrates fermionic behavior, characterized by mutual intrinsic repulsive interaction among particles. This is illustrated by the negative curvature seen in Fig. \ref{R z mb a kam}, a signature of the repulsive interactions typical of fermionic systems.  The observation of this negative curvature corroborates the fermionic nature of the particles within this parameter range.

For \(1 < \alpha \le 1.5\), particles display bosonic properties characterized by a tendency for mutual attraction. This is evident from the positive curvature depicted in Fig. \ref{R z mb a bish}, indicating the presence of bosonic traits. In this case, a critical point \(z_{c}^{\alpha}\) is identifiable, at which the thermodynamic curvature exhibits a divergence. This divergence indicates a phase transition analogous to BEC ~\cite{zare2012condensation, mirza2011condensation}. This critical point is a significant aspect of the MLMB distribution, as it marks the threshold at which collective bosonic behavior becomes evident, leading to the condensation phenomenon.

\section{TRANSITION TEMPERATURES} 

\label{}

\begin{figure}[t]
    \includegraphics[scale=0.5]{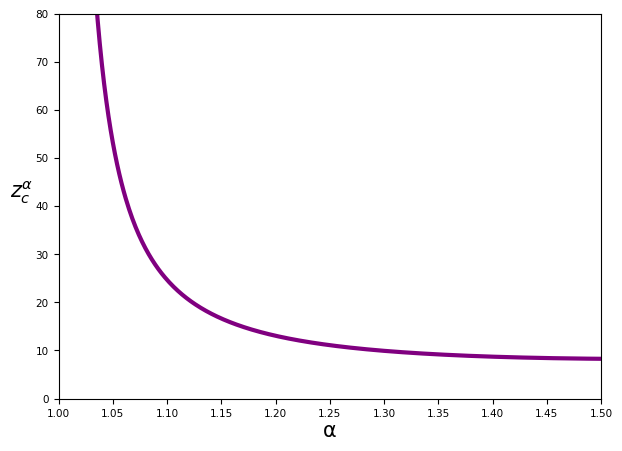}

    \captionsetup{justification=raggedright, singlelinecheck=false} % Left-align caption
    \caption{%
        \(z_{c}^{\alpha}\) values of an MLMB distribution as a function of \(\alpha\), plotted for the range \(1 < \alpha \leq 1.5\).
    }
    \label{z--a ---r}
\end{figure}

In the preceding section, it was noted that the thermodynamic curvature of MLMB distribution diverges at the point \(z_{c}^{\alpha}\) for \(1 < \alpha \leq 1.5\).  Fig. \ref{z--a ---r} depicts the relationship between \(z_{c}^{\alpha}\) and \(\alpha\). Importantly, the value of \(z_{c}^{\alpha}\) is obtained from the divergence of the thermodynamic curvature. Comparison of Figs. (\ref{z--a ---r}) and (\ref{z_max_be_ezay_alpha_ha_plot})  indicates that $z_{\mathrm{th}}^{\alpha}$ , which represents the maximum permissible fugacity for each \(\alpha\) coincide with the critical values of fugacity;  \(z_{c}^{\alpha}\). This consistency is not a coincidence; rather, it emphasizes that \( z_{\mathrm{th}}^{\alpha} = z_{c}^{\alpha} \) indicates the critical point at which the thermodynamic functions display specific behaviors.

Moreover, determining the critical temperature \(T_{c}^{\alpha}\) corresponding to \(z_{c}^{\alpha}\) for each \(\alpha\) is essential for a comprehensive understanding of the phase transition properties. The critical temperature \(T_{c}^{\alpha}\) depends on the parameter \(\alpha\) and must be calculated under the specific conditions of the system. When the number of particles \(N\) is held constant, \(T_{c}^{\alpha}\) can be computed using Eq. (\ref{uN mlmb}) as follows:
\begin{equation}
T_{c}^{\alpha} = \frac{h^{2}}{2\pi m k_{B}} \left(\frac{N}{V \mathcal{H}_{\frac{1}{2},\alpha}(z_{c}^{\alpha})}\right)^{\frac{2}{3}}
\label{ttt zzz}
\end{equation}

To gain a more profound understanding of the thermodynamic behavior within the range of \(1 < \alpha \le 1.5\), we analyzed the variation of the thermodynamic curvature as a function of \(T/T_{c}^{\alpha}\), as shown in Fig. (\ref{R_TbarTc.png}). Here, \(T_{c}^{\alpha}\) represents the critical temperature corresponding to \(z_{c}^{\alpha}\).

As explained earlier, the calculation of the thermodynamic curvature \(R\) from Eq. (\ref{rrr}) requires considering a system with two thermodynamic degrees of freedom, denoted by \(\beta\) and \(\gamma\). In particular, for the case that \( T_c^{\alpha} < T \) and \(z = z_c^{\alpha} \) (a fixed value), the parameter \(\gamma\) remains constant. As a result, the system's degrees of freedom are reduced from two to one, making the system's behavior solely dependent on the \(\beta\) variable. In the case of a one-dimensional system characterized by a flat thermodynamic manifold, the thermodynamic curvature \(R\) reduces to zero. Therefore, The value of \(R\) is zero when \(T_c^{\alpha} > T\). However, some researchers argue that applying Riemannian geometry in a one-dimensional setting is fundamentally inappropriate. Consequently, calculations in regions where the relative temperature is less than one are considered invalid, as the thermodynamic parameter space effectively collapses to a single dimension in this regime.

Conversely, when \( T_c^{\alpha} > T \), it is clear that as \(T\) approaches \(T_c^{\alpha}\), corresponding to \(z = z_c^{\alpha } \), \(R\) shows a clear tendency to diverge. This divergence represents an axial transition point that represents a critical point where significant changes in thermodynamic properties appear.

\section{the heat capacity of MLMB Distribution } \label{}

\begin{figure}[t]
    \flushleft % Left align the figure
    \includegraphics[scale=0.5]{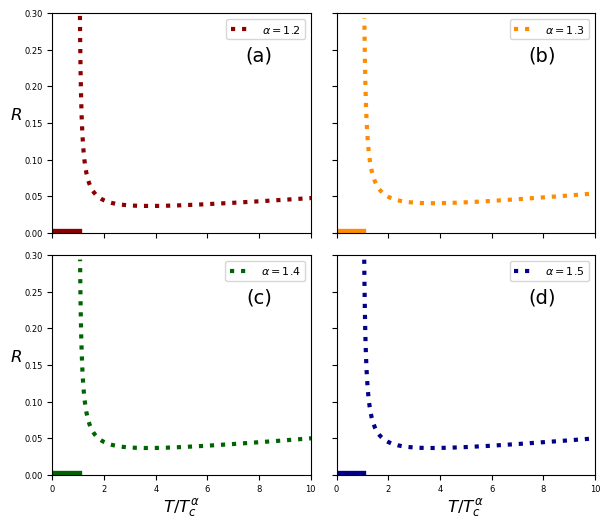}
    
    \captionsetup{justification=raggedright, singlelinecheck=false} % Left-align caption
    \caption{%
        Thermodynamic curvature of an MLMB distribution as a function of \(T/T_{c}^{\alpha}\). 
        For \((a) \, \alpha = 1.2, \, (b) \, \alpha = 1.3, \, (c) \, \alpha = 1.4,\) and \((d) \, \alpha = 1.5\).
    }
    \label{fig:R_TbarTc}
\end{figure}

 In the previous section, we established that the fugacity in the MLMB distribution must have a threshold to maintain a positive distribution function for values of \(1 < \alpha \leq 1.5\). Specifically, the fugacity is restricted to the interval \(0 \leq z \leq z_{c}^{\alpha}\). At \(z = z_{c}^{\alpha}\), the thermodynamic curvature exhibits a divergence, which is also observed in the corresponding critical temperature. 

In this section, we analyze the implications of this divergence on the thermodynamic response functions, with particular emphasis on the heat capacity. The heat capacity \(C_v\) is defined as the partial derivative of the internal energy \(U\) with respect to temperature \(T\) at constant volume, expressed mathematically as:

\begin{equation}
    \frac{C_v}{Nk_{b}} = \left(\frac{\partial U}{\partial T}\right)_v 
    \label{eq:Cv_mb}
\end{equation}

For an ideal classical gas, \(C_v\) solely depends on the degrees of freedom of the gas particles and remains unaffected by external variables like temperature and pressure. In this ideal case, the relationship \(C_v/Nk_{b} =3/2\) is derived.

By utilizing Eqs. (\ref{uNgg mlmb}) and (\ref{eq:Cv_mb}), we extend the classical theory to compute \(C_v\) for MLMB distribution and obtain that

\begin{equation}
 \frac{C_v}{Nk_{b}} = \frac{5 \mathcal{H}_{\frac{3}{2},\alpha}(z)}{2 \mathcal{H}_{\frac{1}{2},\alpha}(z)}-\frac{3 
 \partial_{z}\mathcal{H}_{\frac{1}{2},\alpha}(z)}{2 
 \partial_{z}\mathcal{H}_{\frac{1}{2},\alpha}(z)}.
  \label{e22q:Cv_mb}
\end{equation}

\begin{figure}[t]
    \includegraphics[scale=0.5]{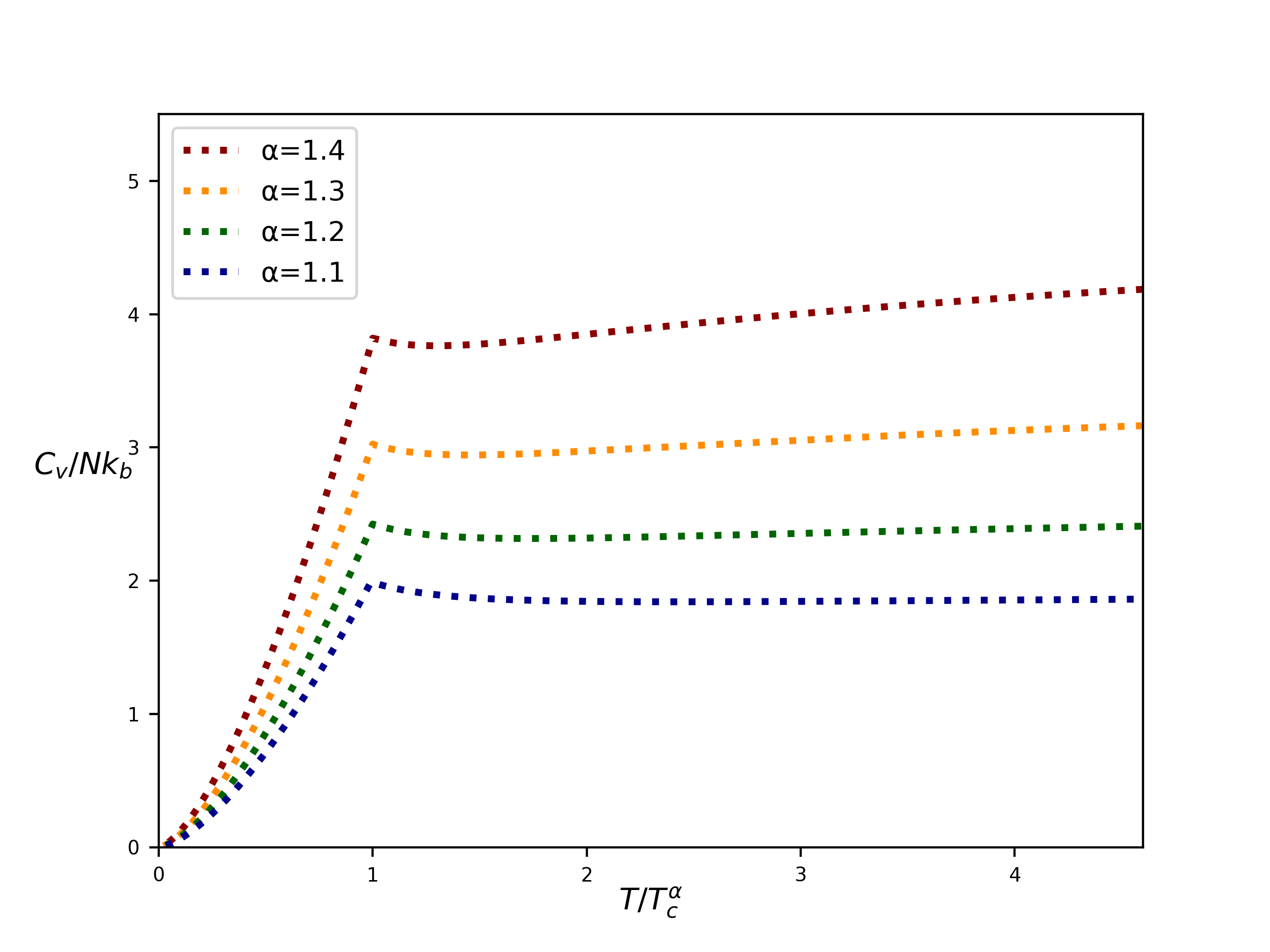}

    \captionsetup{justification=raggedright, singlelinecheck=false} % Left-align caption
    \caption{%
        Heat capacity at fixed volume as a function of temperature, plotted for the range \(1 < \alpha \leq 1.5\). Dashed lines represent the values \(\alpha = 1.1\), \(1.2\), \(1.3\), and \(1.4\).
    }
    \label{cv_alpha_z1}
\end{figure}

As previously mentioned, the parameter \(\alpha\) plays a crucial role in determining the thermodynamic characteristics of the system under consideration. This parameter is instrumental in identifying deviations from classical behavior: when $1<\alpha\le1.5$, the system exhibits boson-like properties, whereas for \( 0 < \alpha < 1 \), it transits to fermion-like behavior. Consequently, a thorough investigation across different \(\alpha\) values is imperative.

Fig. (\ref{cv_alpha_z1}) illustrates the temperature dependence of \( C_v/Nk_{b} \) for the range \(1 < \alpha \leq 1.5\), as derived from Eq. (\ref{e22q:Cv_mb}). In this case, the system demonstrates a phase transition similar to the BE distribution, indicated by a discontinuity in the thermodynamic curve at the critical temperature $T = T_{c}^{\alpha}$. \( T_{c}^{\alpha} \) denotes the temperature associated with \(  z_{c}^{\alpha} \) for each \( \alpha \).Notably, the heat capacity behavior at higher temperatures deviates from the BE distributions. While \( C_v \) for the BE distribution converges to \( C_v/Nk_{b} = 3/2 \), such convergence does not occur within the system for \(1 < \alpha \leq 1.5\). Instead, \( C_v/Nk_{b} \) exhibits a continuous and gradual increase. This difference underscores the unique thermodynamic properties of the system within the specified range of \(\alpha\). Indeed, systems exhibit analogous behavior in the high-temperature limit under classical and quantum statistics. However, when analyzed using the MLMB distribution, the system retains its unique characteristics even at elevated temperatures.

\begin{figure}[t]
    \includegraphics[scale=0.5]{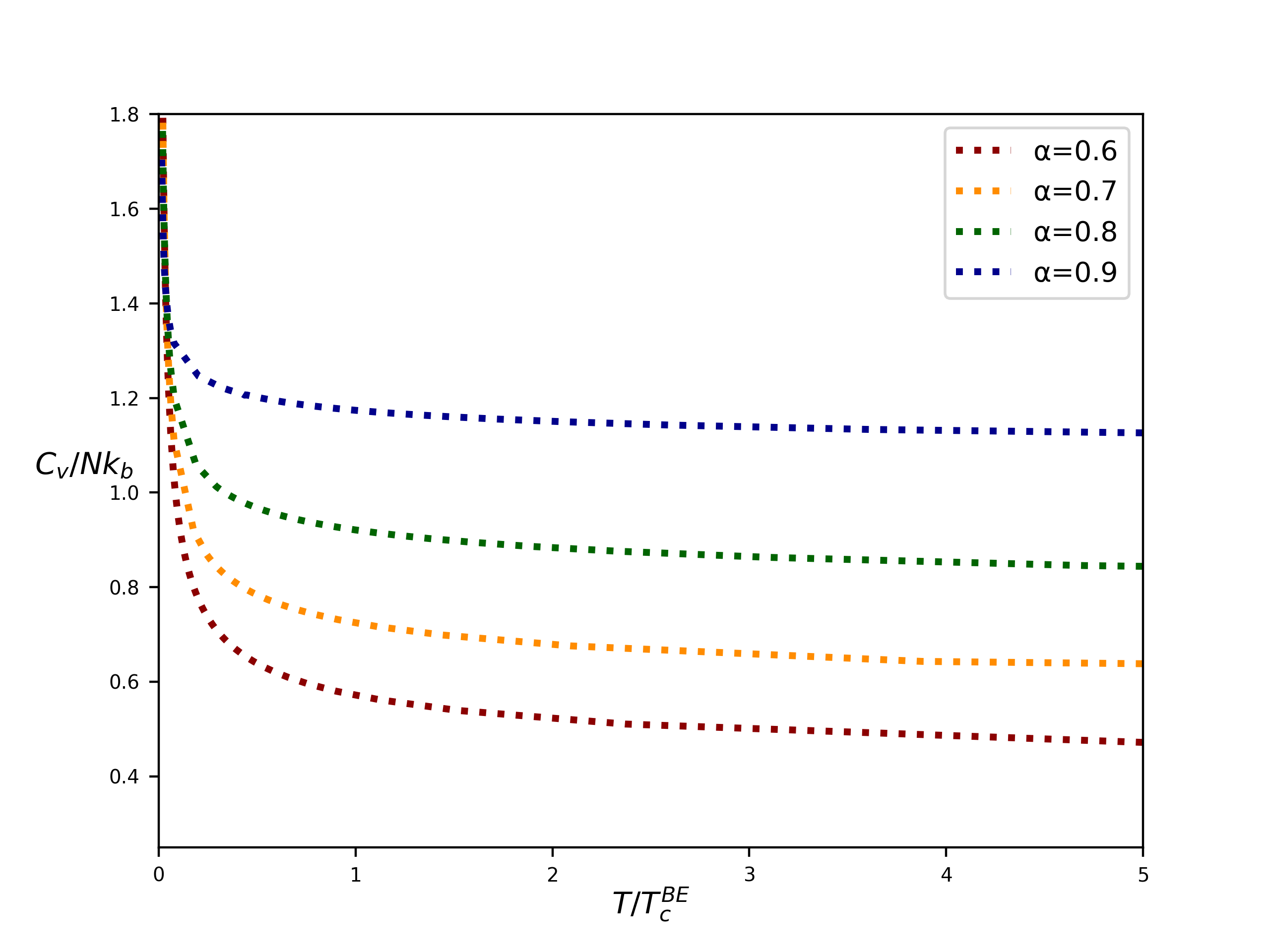}

    \captionsetup{justification=raggedright, singlelinecheck=false} % Left-align caption
    \caption{%
        Heat capacity at fixed volume as a function of temperature, plotted for the range  $1<\alpha<0$. Dashed lines represent the   $\alpha=0.6, 0.7, 0.8, 0.9$.
    }
    \label{cv alpha z2}
\end{figure}

Fig. (\ref{cv alpha z2}) shows the behavior of heat capacity with respect to temperature for different values of \( 0 < \alpha < 1 \), as derived from Eq. (\ref{e22q:Cv_mb}). The temperature scale employed in this graph is normalized based on the critical temperature of the BE distribution $(T_{c}^{BE})$.  The \(  C_v/Nk_{b} \) diverges to positive infinity at lower temperatures.  With an increase in temperature, \(  C_v/Nk_{b} \), gradually and continuously decreases towards zero, never reaching zero.

 \section{Mittag-Leffler Distribution: Evidence, Applications, and Anomaly Analysis}

The Mittag-Leffler (ML) function, a fundamental component of fractional calculus, has been widely applied across various domains of physics and interdisciplinary sciences. As highlighted by Haubold, Mathai, and Saxena (2011), it plays a crucial role in relaxation processes, anomalous diffusion, and fractional kinetic equations, bridging mathematical theory with statistical physics and engineering~\cite{haubold2011mittag}. In astrophysics, José et al. (2009) demonstrated its utility in modeling stochastic processes and time series data, such as solar neutrino flux and cosmic ray events, where its heavy-tailed nature captures long-memory effects~\cite{jose2009generalized}. Arrigo and Durastante (2021) further extended its applicability to network science, where ML functions characterize complex dynamics, including fractional diffusion on graphs, aiding in the study of information propagation and resilience in large-scale networks~\cite{arrigo2021mittag}. Additionally, dos Santos (2020) integrated ML functions into superstatistics, providing a robust framework for analyzing systems with multi-scale fluctuations, including turbulent plasmas and non-equilibrium thermodynamics~\cite{dos2020mittag}. These studies collectively underscore the versatility of the ML function in addressing diverse challenges across astrophysics, networked systems, and statistical mechanics.

Given the extensive applications of the ML distribution, we employ it to generalize the Maxwell-Boltzmann (MB) statistics. As illustrated in Fig.~(\ref{cv alpha z2}), a notable consequence of the MLMB distribution is the divergence of heat capacity in a fermion-like regime. While this may initially seem nonphysical, similar divergences have been observed in various physical systems, particularly those exhibiting non-trivial interactions, quantum fluctuations, or topological effects.

In fermionic systems, such as marginal Fermi liquids and certain non-Fermi liquid models, electronic heat capacity follows a power-law behavior and may diverge as \( T \to 0 \) due to deviations from the Landau Fermi-liquid framework~\cite{varma1997non, senthil2004weak, stewart2001non}. Similarly, in low-dimensional Bose-Einstein condensation, quantum fluctuations dominate, leading to heat capacity divergences due to the unique scaling behavior of the density of states~\cite{pathria2011statistical, ketterle1996bose}. Quantum critical systems, which undergo phase transitions at absolute zero temperature, exhibit non-analytic behavior in heat capacity as they are governed by quantum fluctuations rather than thermal fluctuations~\cite{sachdev1999quantum, gegenwart2008quantum}.

Topological excitations, such as vortices, monopoles, and domain walls, also influence heat capacity behavior. In spin ice materials, for example, emergent magnetic monopoles contribute to residual entropy and non-vanishing heat capacity even as \( T \to 0 \)~\cite{mermin1979topological, castelnovo2008magnetic}. Strongly correlated or topologically ordered systems host exotic quasiparticles like Majorana fermions, spinons, and anyons, whose gapless spectra result in power-law or logarithmic divergences in heat capacity at low temperatures~\cite{kitaev2006anyons, balents2010spin}.

In black hole thermodynamics, certain black hole solutions exhibit divergent heat capacity as their temperature approaches zero. This occurs in extremal limits of charged or rotating black holes, where quantum effects modify classical thermodynamics. For instance, Reissner-Nordström-AdS black holes in a canonical ensemble, as well as those governed by higher-curvature Einstein-Gauss-Bonnet gravity, display heat capacity divergences near extremality~\cite{chamblin1999charged, cai1999critical}. Similarly, black holes with hyperscaling-violating metrics in Einstein-Maxwell-dilaton theories exhibit non-trivial scaling behavior near extremality, leading to divergent heat capacity due to the interplay between the dilaton field and charge, which significantly alters the near-horizon geometry~\cite{gouteraux2013quantum}.

These examples demonstrate that divergent heat capacity behavior is a well-established feature of many physical systems, reinforcing the potential relevance of the MLMB distribution in capturing anomalous thermodynamic properties. Exploring heat capacity anomalies in the discussed examples through the Mittag-Leffler distribution presents a promising direction for future research.

\section{CONCLUSION}\label{}

In this work, we introduced a generalized form of the classical Maxwell-Boltzmann distribution by employing the Mittag-Leffler function within the framework of superstatistics. This generalization enabled us to explore the statistical interactions present in various systems, where the parameter $\alpha$ plays a crucial role in modulating these interactions. Specifically, we demonstrated that for $\alpha = 1$, the system reverts to the classical Maxwell-Boltzmann distribution with no statistical interactions, while values of $\alpha$ in the range $0<\alpha < 1$ and $\alpha > 1$  introduce a non-trivial statistical interaction. Our analysis reveals that the thermodynamic curvature, which is flat for ideal classical gases, acquires non-zero values for generalized distributions, indicating the presence of intrinsic attractive and repulsive interactions similar to those observed in quantum Bose and Fermi gases. Importantly, the $\alpha$ parameter emerges as a robust tool for characterizing thermodynamic curvature, providing researchers with a quantitative measure for evaluating and comparing distributional characteristics. Thermodynamic curvature's dependence on the parameter $\alpha$ highlights how adaptable our approach is to different physical scenarios.

Notably, the investigation of intermediate statistics related to the emergence of bosonic and fermionic behaviors reveals substantial differences. In Haldane statistics, it has been shown that when the fugacity falls below a certain critical threshold, the system can exhibit bosonic behavior. Nevertheless, across most of the physical domain and for the majority of fractional parameter values, the system predominantly exhibits fermionic behavior ~\cite{mirza2010thermodynamic}. In the case of Tsallis statistics, the introduction of the generalized parameter preserves the inherent bosonic and fermionic characteristics without modification  ~\cite{mohammadzadeh2016perturbative,adli2019nonperturbative}. In Kaniadakis statistics, fermionic behavior remains unchanged when exponential functions are substituted with generalized Kaniadakis functions; however, bosonic systems may exhibit fermion-like characteristics within certain temperature ranges ~\cite{mehri2020thermodynamic}. Recent studies on unified quantum statistics have demonstrated that for specific values of the generalized parameter, a distinct temperature threshold can be determined, above which the system exhibits fermionic behavior, while below it, bosonic characteristics prevail  ~\cite{esmaili2024quantum}. The statistical framework we have introduced is uniquely capable of explicitly differentiating between bosonic and fermionic categories solely based on the generalized parameter, independent of temperature and fugacity.

The implications of this work are significant for the study of generalized thermodynamics, where non-extensive systems or systems with long-range interactions can be modeled more accurately. Our approach provides a flexible framework to analyze various statistical ensembles and can be extended to study systems with more complex interactions. Future work may explore the application of this generalized distribution to real-world systems, such as non-equilibrium statistical mechanics, astrophysical phenomena, or condensed matter systems where deviations from classical behavior are prominent.

In conclusion, the Mittag-Leffler Maxwell-Boltzmann distribution offers a powerful tool for extending classical statistical mechanics into new regimes, opening avenues for further research into both theoretical and applied physics.

%%%%%%%%%%%%%%%%%%%%%%%%%%%%%%%%%%%%%%%%%%%%%%%%%%%%%%%%%%%%%%%%%%%%%%%%%%%%%%%%%%%%%%%%%%%%%%%%%%%%%%%%%%%%%%%%%%%%%%%%%

\section{appendix}\label{appendix}

\textbf{Derivative of the Mittag-Leffler Function}

The derivative of the single-parameter Mittag-Leffler function with respect to \(z\) is given by:

\begin{equation}
 \frac{\partial}{\partial z} E_{\alpha}(z) = \frac{E_{\alpha, \alpha}(z)}{\alpha},
 \label{mlf_derivative}
\end{equation}

where \( E_{\alpha, \alpha}(z) \) denotes the two-parameter ML function with \( \lambda = \alpha \). Notably, when the second parameter of the two-parameter ML function coincides with the first parameter \( E_{\alpha, \alpha}(z) \), the two-parameter ML function can be expressed as the derivative of the single-parameter ML function.

Next, consider a more general expression involving the function \( \mathcal{H}_{n,\alpha}(z) \) which can be defined by the integral:

\begin{equation}
\mathcal{H}_{n,\alpha}(z) = \int_{0}^{\infty} \frac{x^{n}}{E_{\alpha}(x - \ln z)} \  dx.
\label{H_function}
\end{equation}

Taking the derivative \( \mathcal{H}_{n,\alpha}(z) \) with respect to \(z \) gives the result:

\begin{align}
\frac{\partial \mathcal{H}_{n,\alpha}(z)}{\partial z} &= \frac{\partial}{\partial z}\left(\int_{0}^{\infty} \frac{x^{n}}{E_{\alpha}(x - \ln z)} \, dx \right) \nonumber \\
&= \int_{0}^{\infty} \frac{x^{n}  E_{\alpha, \alpha}(x - \ln z)}{\alpha z \left(E_{\alpha}(x - \ln z)\right)^{2}} \, dx.
\label{H_derivative}
\end{align}

This derivative expresses the complex interplay between the ML functions, revealing the intricate nature of their functional forms. This expression plays a role in calculating the metric elements of thermodynamic curvature.
%%%%%%%%%%%%%%%%%%%%%%%%%%

\bibliographystyle{apsrev}
\bibliography{re}

\end{document}